\documentclass[10pt]{article}    

\usepackage{amssymb,latexsym,amsmath} 
\usepackage[mathscr]{eucal}  
\newfont{\sffl}{msbm10 at 16pt} 
\newfont{\sff}{msbm10 at 10pt}

\begin{document}           
\title{\vskip -.75in
Closed-form Dirichlet integral harmonic interpolation-fits for\\
   real n-dimensional and complex half-space: DIDACKS III
\thanks{\small{Approved for public  release.}}}

\author{Alan Rufty\\         
\\
P.O. Box 711\\
Dahlgren, VA. 22448}
\maketitle                 

\newcommand{\KD}{K_{\text{D}}}

\begin{abstract}
  This article gives a brief overview of a ``fundamental solution'' based energy-norm harmonic interpolation approach for two half-space settings of interest: the upper-half $\mathbb{R}^n$ plane, where fundamental solutions satisfy Laplace's equation, and the upper-half complex plane, where simple poles can be regarded as the fundamental kernels of interest (versus fundamental solutions).  It is also pointed out that the approach can handle higher-order pole fits, as well as logarithmic source fits, in the complex setting and that it can handle higher-order multipole fits in the general real $\mathbb{R}^n$ setting.  Higher-order multipoles in the real $\mathbb{R}^n$ half-space setting are of particular interest since fits based on a commonly used type of  radial-basis functions, which are known as inverse multiquadrics, can be reinterpreted as multipole based interpolations with the formalism presented here; moreover, this reinterpretation has certain significant previously unrecognized theoretical implications. 
 
  Although this article can be read independently from the articles that preceded it in the series, a brief look at the content of these preceeding articles is useful here.
In the first article a simple interpolation approach that satisfies a minimum energy-norm condition was presented for representing analytic functions in the exterior of a complex unit disk by a linear combination of simple poles located in the disk's interior.  Corresponding interpolation techniques for the exterior of real unit disks and for the interior of real and complex unit disks were also presented.  Relationships of this approach to Bergman  and Szeg\H{o} kernel interpolation theory were also addressed.  The second article gave a corresponding technique for using reciprocal-distance fundamental solution basis functions to perform harmonic interpolation over the interior of an $\mathbb{R}^3$ sphere.  This $\mathbb{R}^3$ technique is based on closed-form weighted Dirichlet integral inner products, which means that when source positions are specified in the exterior of the harmonic sphere a closeted-form linear equation set results for the source parameters and that the resulting solution minimizes the weighted field energy of the modeling errors.  It was also pointed out in this second article that a similar formalism exists for the exterior of sphere and for $\mathbb{R}^3$ half-space, where the Dirichlet integral itself is directly minimized.  The primary associated functional analysis setting for all of these developments is labeled Dirichlet integral dual-access collocation-kernel space (DIDACKS).  Both previous articles showed that the basic DIDACKS approach can be easily extended to handle higher-order (multi-) poles.  Finally, this second article also presented a proof that not only is the (weighted) Dirichlet integral norm of the modeling difference minimized, but also that of all candidate interpolating functions, the DIDACKS solution is the one with a minimum norm.  

   This article  derives analogous results for complex half-space and then generalizes the $\mathbb{R}^3$ half-space results to $\mathbb{R}^n$ half-space.   This means that closed-from expressions are presented that yield interpolating fits, which directly minimize Dirichlet integral expressions in either $\mathbb{C}$ and $\mathbb{R}^3$ half-space.  Moreover, the DIDACKS solution satisfies a second minimization condition, which states that any interpolating function that matches the prescribed point data values must have a field energy content that is at least as large as that of the DIDACKS solution.  This also means  that when the above reinterpretation of inverse multiquadric radial-basis function fits as multipole DIDACKS fits can be carried out, these radial-basis function fits also simultaneously satisfy two energy-norm minimization conditions, which in itself implies a certain level of numerical robustness. 

    Finally, this article also shows that the same sort of development can be carried out for $\mathbb{R}^n$ half-space surface integral norms where the square integral of the modeling error itself is directly minimized over the half-space boundary (hyper) plane.  This development allows for a reinterpretation of radial basis fits similar to that just described; moreover, it also yields a mathematically motivated approach to the problem of half-space downward continuation.
\end{abstract}

\newcommand{\SubSec}[1]{

\vskip .18in
\noindent
\underline{{#1}}
\vskip .08in

}
\newcommand{\ls}{\vphantom{\big)}} 
\newcommand{\lsm}{\!\vphantom{\big)}} 

\newcommand{\eq}{:=}

\newcommand{\smallindex}[1]{\text{\raisebox {1.5pt} {${}_{#1}$}}}

\newcommand{\mLarge}[1]{\text{\begin{Large} $#1$ \end{Large}}}
\newcommand{\mSmall}[1]{\text{\begin{footnotesize} $#1$ \end{footnotesize}}}

\newcommand{\Blbrac}{\rlap{\bigg{\lceil}}\bigg{\lfloor}} 
\newcommand{\Brbrac}{\bigg{\rbrack}} 

\newcommand{\Dr}{\mathscr{D}_r}
\newcommand{\R}[1]{${\mbox{\sff R}}^{#1}$} 
\newcommand{\mR}[1]{{\mbox{\sff R}}^{#1}}
\newcommand{\RR}{${\mbox{\sff R}}$}
\newcommand{\mRR}{{\mbox{\sff R}}}
\newcommand{\C}{${\mbox{\sff C}}$}
\newcommand{\mC}{{\mbox{\sff C}}}
\newcommand{\D}{{\text{D}}}
\newcommand{\sps}{0}

\newcommand{\smallindexes}[1]{\text{\raisebox {.5pt} {${}_{#1}$}}}

\newcommand{\hc}{\text{H}_{{\text{\raisebox {1.5pt} {${}_{\mspace{-1.3mu}\perp}$}}}}^{\mspace{1.5mu}\text{\raisebox {1pt} {$c$}}}}
\newcommand{\Hc}{\text{H}^c}

\newcommand{\hn}{\text{H}_{{\text{\raisebox {1.5pt} {${}_{\mspace{-1mu}\perp}$}}}}^{\mspace{1mu}\text{\raisebox {1pt} {$n$}}}}
\newcommand{\Hn}{\text{H}^n}
\newcommand{\Harg}[1]{\text{H}_{{\text{\raisebox {1.5pt} {${}_{\mspace{-1.1mu}\perp}$}}}}^{#1}}

\vskip .05in
\noindent
\begin{itemize}
\item[\ \ ] \small{\textbf{Key words:} {complex half-space, interpolation, Dirichlet form, radial basis function, Laplace's equation,\\  \phantom{Key words. L} complex poles , inverse problem, Dirichlet form, point sources, multipole, potential theory,\\  \phantom{Key words. L}} fundamental solutions}
\item[\ \ ] \small{\textbf{AMS subject classification (2000):} {Primary 31Bxx. Secondary 86A22, 35J99, 65D05}}
\end{itemize}

\renewcommand {\baselinestretch}{1.35} 
 
\section{Introduction}\label{S:intro}

  This article develops a simple closed-form interpolation approach that minimizes the associated field energy over the half-space region of interest by using a linear combination of poles in $\mathbb{C}$, or fundamental solution basis functions in $\mathbb{R}^n$, as interpolating functions.  This means that in both settings, the resulting formalism can simultaneously be considered as an interpolation technique and as an approximation technique. 

   First consider the approximation aspects of the resulting formalism.  In the complex setting, where $z = x + iy$, the formalism approximates a given function $f(z)$ that is analytic in the upper-half plane $\Hc \eq \{ z \in  \mC\mid\, y \geq 0 \}$ by a finite set of spatially distributed simple poles in the lower-half plane $\hc \eq \{ z \in  \mC\mid\,  y < 0 \}$ so that the specific approximating (and interpolating) form becomes
\begin{equation}\label{E:fitform}
 \varphi = \sum\limits_{k=1}^{N_k} \frac{{\mu}_k}{z - z_k}
\end{equation}
 with $\mu_k \in \mathbb{C}$ and $z_k \in \hc$. 

  Likewise, in the real $\mathbb{R}^n$ setting, where $\vec{X} = (x_1,\,x_2,\,x_3,\, \ldots,\,x_n)^T$, if $f(\vec{X})$ is harmonic in the $n$-dimensional upper-half plane
$\Hn \eq \{ \vec{X} \in  \mathbb{R}^n \mid\, x_n \geq 0 \}$, then it is to be approximated by a finite set of spatially distributed fundamental solutions that have sources in the lower half-space $\hn \eq \{ \vec{X} \in  \mathbb{R}^n \mid\, x_n < 0 \}$, so that the approximating (and interpolating) form becomes
\begin{equation}\label{E:fitform2}
 \varphi = \sum\limits_{k=1}^{N_k}\ \frac{{\mu}_k\ \ }{|\vec{X} - \vec{X}'_k|^{n-2}}
\end{equation}
with  $\mu_k \in \mathbb{R}$ and  $\vec{X}'_k \in \hn$.  (Other various notational conventions were explained in \cite{DIDACKSI,DIDACKSII}.)

   For a given $f$, when the values of the source locations are specified, the usual approximation strategy of introducing a norm and then minimizing the associated cost function
\begin{equation}\label{E:PHI}
 \Phi \eq \|f - \varphi\|^2 = \|f\|^2 - 2(f,\,\varphi) + \|\varphi\|^2
\end{equation}
can then be used to determine the unknown source coefficients.  In order to consider not only fits like (\ref{E:fitform}) and (\ref{E:fitform2}), but other general forms as well, suppose that $\varphi$ is a linear combination of general basis functions:
\begin{equation}\label{E:LLSQbasis}
\varphi(z) = \sum\limits_{k=1}^{N_k} {\mu}_k{\mathcal{B}}_k\ .
\end{equation}
 For the real setting, the values of ${\mu}_k$ that minimize $\Phi = \Phi({\mu}_k)$ can be found simply by solving the linear equation set that results from setting the partials of $\Phi$ with respect to ${\mu}_{k'}$ to zero (and diving by two): 
\begin{equation}\label{E:LLSQ4}
\sum\limits_{k=1}^{N_k} ({\mathcal{B}}_{k'},\,{\mathcal{B}}_k){\mu}_k = ({\mathcal{B}}_{k'},\,f) \ ,
\end{equation}
which can be written more compactly as
\begin{equation}\label{E:TmuA}   
\boxed{\,\mathbf{T}\,\mathbf{\mu} = \mathbf{A}\,} 
\end{equation}
 where $\mathbf{T}$ denotes the matrix whose elements are $T_{k',\,k} \eq ({\mathcal{B}}_{k'},\,{\mathcal{B}}_k)$, $\mathbf{A}$ the vector whose element are  $A_{k} \eq ({\mathcal{B}}_{k},\,f)$ and $\mathbf{\mu}$ the vector whose elements are ${\mu}_k$.

  For the complex setting, when complex conjugation is applied to the first factor in an inner product, rather than the second, it is said to be in left conjugate form (LCF) \cite{DIDACKSI} and it is a simple matter to show that (\ref{E:TmuA}) also specifies the solution that minimizes the cost function $\Phi$ specified by (\ref{E:PHI}) \cite{DIDACKSI}.  In the complex setting $\mathbf{T}$ is obviously Hermitian.

  In the real $\mathbb{R}^n$ setting for $n \geq 2$, it is assumed that $f$ and $g$ are harmonic over $\Omega$, which may be an unbounded domain, and that the inner products are specified in terms of Dirichlet integrals so that $|f|$ and $|g|$ must fall off to zero sufficiently fast as $|\vec{X}| \rightarrow \infty$ \cite{DIDACKSI,DIDACKSII}.  Recall that the Dirichlet integral is usually denoted by $\text{D}[f,\,g] = \int_\Omega {\nabla} f\cdot{\nabla} g \,\,dV$ and that this notation was generalized in an obvious fashion in \cite{DIDACKSII} to include a weighting function $\mu = \mu(\vec{X}) > 0$ and to indicate the domain $\Omega$ of interest:
\begin{equation}\label{E:DnInt}
\text{D}[f,\,g,\,\mu,\,\Omega]\, \eq  \,\int\limits_\Omega \mu\,{\nabla} f\cdot{\nabla} g \,\,dV\ . 
\end{equation}
While only the case $\mu = 1$ is of interest for half-space and $\text{D}[f,\,g,\,1,\,\Omega] = \text{D}[f,\,g]$, this generalized Dirichlet integral notation will still often be used here in order to indicate the domain under consideration.  Thus, for $\mathbb{R}^n$ half-space the primary inner product under consideration will be
\begin{equation}\label{E:DnIP}
(f,\,g){\ls}_{{\text{D}}/\Hn}\, \eq \,\text{D}[f,\,g,\,1,\,\Hn]\ . 
\end{equation}

   Whereas these same approximation considerations apply to any appropriate set of basis functions $\{{\mathcal{B}}_k\}_{k=1}^{N_k}$ defined over the region of interest $\Omega\,$; the interpolation attributes hold only for $\Omega$'s with certain specific geometries of interest and these attributes result from the interaction of the associated norms defined over $\Omega$ with these particular basis functions.  Specifically for an appropriate $\mathbb{R}^n$ inner product environment, where $\Omega \subset \mathbb{R}^n$,  if the basis functions ${\mathcal{B}}_{k}(\vec{X})$ have a point reproducing or replication property, then the inner-products of interest involving ${\mathcal{B}}_{k}$ can be written in the form
\begin{equation}\label{E:Bh}
({\mathcal{B}}_{k},\,f) = h({\vec{P}}_k)f({\vec{P}}_k)\,\,,
\end{equation}
where ${\vec{P}}_k \in \Omega$ and
$h({\vec{P}}_k) \neq 0$ is a fixed function \cite{DIDACKSII}.  Corresponding point reproducing or replication relationships hold for the complex setting, but for the purposes of discussion just the $\mathbb{R}^n$ setting is considered for now.  [Also although ignored here for simplicity, in general, the possibility of a linear differential operator is also considered on the right hand side (RHS) of (\ref{E:Bh})---see, for example, (\ref{E:CollPropD}).]

Observe that there are two general situations of interest here.  The first is when ${\mathcal{B}}_{k}(\vec{X})$ can naturally be written as a function of two arguments that are both in $\Omega$, say ${\mathcal{B}}_{k} = {\mathcal{B}}(\vec{X},\,{\vec{P}}_k)$ where $\vec{X}$ and ${\vec{P}}_k \in \Omega\,$, and the second is when ${\mathcal{B}}_{k}(\vec{X})$ can naturally be written as a function of two arguments that are in two different domains, say ${\mathcal{B}}_{k} = {\mathcal{B}}(\vec{X},\,{\vec{X}}'_k)$ where $\vec{X} \in {\Omega}$ and ${\vec{X}}'_k \in {\Omega}'$, which is the compliment of ${\Omega}$.  This second situation is of special interest in what follows and, further, ${\vec{X}}'_k = {\vec{X}}'_k(\vec{P}_k)$ and ${\vec{P}}_k = \vec{P}_k({\vec{X}}'_k)$ hold for all the particular cases studied here, where ${\vec{P}}_k \in \Omega$. 
[Observe that for reproducing kernel Hilbert spaces $h({\vec{P}}_k) = 1$ and $K(\vec{X},\,{\vec{P}}_k) \eq  {\mathcal{B}}_{k}(\vec{X}) = {\mathcal{B}}(\vec{X},\,{\vec{P}}_k) = K({\vec{P}}_k,\,\vec{X})$, where $\vec{X}$ and ${\vec{P}}_k \in \Omega$ and  $K(\vec{X},\,{\vec{P}}_k)$ is called the reproducing kernel.]  
When a set of basis functions satisfy a point reproducing or replication property and  $\mathbf{T}^{-1}$ exists then one can show that solutions specified by (\ref{E:TmuA}) have the interpolation property:
\begin{equation}\label{E:Pcoll}
\varphi({\vec{P}}_k) = f({\vec{P}}_k)\ \ \text{for}\ \ k = 1,\,2,\,3,\,\ldots,\,N_k\,\,,
\end{equation}
but how can one show in general that $|{T}| \neq 0$?  

  For the regions of interest, $\Omega$, it was shown in \cite{DIDACKSII} that harmonic functions have a strong enough property to guarantee that all sets of basis functions satisfying a point reproducing or replication property of the form (\ref{E:Bh}) are linearly independent.  A set of functions with this property is labeled uniformly pointwise independent and it can be defined as follows:

\vskip 7pt

\noindent
{\bf{Definition \ref{S:intro}.1}}\ \ 
A set $\mathscr{F}$ of scalar valued functions is said to be \emph{uniformly pointwise independent} over a region $\Omega \subset \mathbb{R}^n$ if for every finite $N$ the following holds: For any bounded set of $N$ arbritrary points ${\vec{X}}_j \in \Omega$ and arbritrary bounded constants $C_j$, there always exists a function $f \in \mathscr{F}$ such that $f({\vec{X}}_j) = C_j$ for $j = 1,\,2,\,3,\,\ldots,\,N$.  
\vskip 7pt

  As shown in \cite{DIDACKSII}, when a point reproducing or replication property of the form (\ref{E:Bh}) holds, this condition is sufficient to guarantee that the following two distinct minimum norm conditions are simultaneously satisfied:
\begin{itemize}
\item
 First,
\begin{equation}\label{E:Pcoll2}
\|f - \varphi\|\ \ \text{is minimized} 
\end{equation}
directly by (\ref{E:TmuA}). 
\item
 Second, of all those functions $g$ that satisfy
\begin{equation*}\notag
g({\vec{P}}_k) = f({\vec{P}}_k)\ \ \text{for}\ \ k = 1,\,2,\,3,\,\ldots,\,N_k\,\,,
\end{equation*}
 $\varphi$, with coefficients fixed by (\ref{E:TmuA}), has the least norm:
 \begin{equation}\label{E:Pcoll3}
  \|g\| \geq \|\varphi\|\ .
 \end{equation}
\end{itemize}
(Trivially, if $a^2 > b^2$ then $|a| > |b|$ and conversely, so that $\|f - \varphi\|^2$ and $\|\varphi\|^2$ are also concurrently minimized.)  For the various DIDACKS half-space settings, since the norms of interest are energy based norms, these two minimization criteria imply the associated solutions are minimum field energy solutions, which has significant physical implications in-and-of itself. 

  Definition \ref{S:intro}.1, as well as the associated discussion centering on (\ref{E:Bh})
through (\ref{E:Pcoll3}), extends in a obvious way to the complex setting so this extension is automatically presupposed in the sequel.

  Having considered the general framework, consider specific examples of inner products and kernels of interest from \cite{DIDACKSI,DIDACKSII}, which display a point replication property:
\begin{enumerate}
\item
\textbf{Standard Integral Norm Setting for the Exterior of a Complex Unit Disk:}\\
Here it is assumed that $|z| \geq 1$ and that $f$ and $g$ have a power series representation of the form:
\begin{equation}\label{E:series1}
 f(z) = \sum\limits_{j=1}^{\infty} \frac{a_j}{z^{j}}\,,
\end{equation}
with $a_j \in \mC$; however, the $a_j$'s need not be known explicitly.  Then the kernel $1/(z - z_k)$ for $0 < |z_k| < 1$ has the point replicating property \cite{DIDACKSI}
\begin{equation}\label{E:sigg3}
 (\,(z - z_k)^{-1},\,f){\ls}_{\sigma} = p_k^{*}\,f(p^{*}_k)\ ,
\end{equation}
where $p_k = 1/z_k$ and the associated standard integral inner product is defined by
\begin{equation}\label{E:sigg1}
(f,\,g){\ls}_{\sigma} \eq  \frac1{2\pi} \int\limits_{\theta=0}^{2\pi} [(f(z))^*\,g(z)]\Big|_{r=1} d\,\theta\  
\end{equation}
in the usual polar coordinates (i.e., $z = re^{i\theta}$, $z_k = r_ke^{{i\theta}_k}$).
\item
\textbf{Dirichlet Integral Norm Setting for the Exterior of a Complex Unit Disk:}\\
 Here it is again assumed that $|z| \geq 1 > |z_k| > 0$ and that, at least in principle, $f(z)$ and $g(z)$ have the series form (\ref{E:series1}).  Then it is easy to show that the kernel $1/(z - z_k)$ has the point replicating property \cite{DIDACKSII}
\begin{equation}\label{E:CollPropD}
(\,(z - z_k)^{-1},\,f){\ls}_{D} = -\left(\frac{p_k^2}2\right)^{*}f'(p_k^{*})\ ,
\end{equation}
where again $p_k \eq 1/z_k$ and 
\begin{equation}\label{E:dfdz}
f'(z) \eq \frac{d\,f}{d\,z}\ .
\end{equation}
Here the complex form of the Dirichlet integral inner product is given by
\begin{equation}\label{E:CplxDirInt}
(f,\,g){\ls}_{D} \eq \frac1{2\pi}\int\limits_{r=1\ }^{\,\,\infty}\int\limits_{\theta=0}^{\,\,2\pi} \bigg(\frac{d\,f}{d\,z}\bigg)^{*}\bigg(\frac{d\,g}{d\,z}\bigg)\, r\,d\,r\,d\,\theta\ ,
\end{equation}
where, as above, a factor of $1/2\pi$ has been introduced for convenience. (The relationship of this inner product to the corresponding $\mathbb{R}^2$ Dirichlet integral $\text{D}[f,\,g,\,1,\,\Omega]$, including the factor of $1/2\pi$, was discussed at length in \cite{DIDACKSI}.)
\item
\textbf{Complex Unit Disk Logarithmic Dirichlet Integral Norm Expressions:}\\
  The inner product is the one just studied for the exterior of a complex unit disk and $z_k$ and $p_k$ have the same meaning.  Now, however, one basis function choice to be considered is
\begin{equation}\label{E:psilog}
  \psi_k(z) = \ln\, \frac{z}{(z - z_k)}
\end{equation}
and for this basis function the point replication condition becomes
\begin{equation}\label{E:psiRep}
  (\psi_k,\,f){\ls}_{D} = \frac12\,f(p_k^{*})\ .
\end{equation}

  The basis function choice specified by (\ref{E:psilog}) is only a special case.  In general it is useful to consider the paired point formalism introduced in \cite{DIDACKSI}, which has the associated basis function
\begin{equation}\label{E:xibasis1}
 {\xi}_k(z) \eq \ln\,\frac1{z - z_k} - \ln\,\frac1{z - z'_k}\, = \,\ln\,\left(\frac {z - z'_k}{z - z_k}\right)\ .
\end{equation}
 Here $z'_k$ is called the paired point of $z_k$ and it is an arbritrary complex number, except for the restrictions that $|z'_k| < 1$ and $z'_k \neq z_{k'}$ for $k' = 1,\,2,\,3,\,\ldots,\,N_k $.  [Clearly the choice  $z'_k = 0$ for all $k$ results in (\ref{E:psilog}).]  For a general paired-point basis function
\begin{equation}\label{E:xi}
({\xi}_k,\,f){\ls}_{D} = \frac12\,f(p_k^{*}) - \frac12\,f((p'_k)^{*})\ ,
\end{equation}
where $(p'_k)^{*}\,$ is the conjugate involution of the paired point of $z_k$ and is thus specified by taking the conjugate of $p'_k \eq 1/{z'_k}$.
\item
\textbf{Dirichlet Integral Norm Setting for $\mathbb{R}^3$ Half-space:}\\
Here (\ref{E:fitform2}) becomes
\begin{equation}\label{E:fitformR3}
 \varphi = \sum\limits_{k=1}^{N_k}\ \frac{{\mu}_k}{{\ell}_k}
\end{equation}
where vectors in the compliment of the field region are denoted by primed vectors so that ${\ell}_k \eq |\vec{X} - \vec{X'_k}|$, where $\vec{X}'_k = (x'_k,\,y'_k,\,z'_k)^T$ denotes a typical source point for $k = 1,\,2,\,3,\,\ldots,\,N_k $.  Since  $\vec{X} = (x,\,y,\,z)^T \in \text{H}^3$ and $\vec{X}'_k \in \Harg{3}$, $z \geq 0$ and $z'_k < 0$, so that $1/{\ell}_k^{-1}$ is always bounded for a specified $\vec{X}'_k$.  (Of course, it is also assumed that $\vec{X}'_{k'} \neq \vec{X}'_k$ for all $k' \neq k$.)  The $\text{H}^3$ point replication (or generalized collocation) property for the DIDACKS kernel ${\ell}_k^{-1}$ is simply
\begin{equation}\label{E:rep1}
 \text{D}[f,\,{\ell}_k^{-1},\,1,\,\text{H}^3]\, =\, 2\pi\,f(x'_k,\,y'_k,-z'_k)\ .
\end{equation}
\end{enumerate}

\begin{center} 
 \hfil\\
\textbf{\underline{Higher-order pole and multipole fits}}
 \hfil\\
\end{center}

Other relationships similar to (\ref{E:sigg3}), (\ref{E:CollPropD}), (\ref{E:psiRep}) and (\ref{E:xi}) were given in \cite{DIDACKSI}, including corresponding $\mathbb{R}^2$ point replication conditions.  A discussion of analogous relationships for the interior of unit disks was also included.  Only the complex half-space relationships corresponding to (\ref{E:sigg3}), (\ref{E:CollPropD}), (\ref{E:psiRep}) and (\ref{E:xi}) are derived here since the other corresponding $\mathbb{R}^2$ relationships can be derived by a relatively straightforward repetition of the steps followed in \cite{DIDACKSI}.  Further, in the complex setting, as observed in \cite{DIDACKSI}, higher order poles can be easily incorporated into the fitting function $\varphi$ since
\begin{equation}\label{E:HOP}
 \frac{d^m\ }{d\,z^m} \bigg(\frac1{z - z_k}\bigg) = \frac{(-1)^m\,m!}{(z - z_k)^{m+1}} = -\frac{d^m\ }{d\,z_k^m} \bigg(\frac1{z - z_k}\bigg)\ 
\end{equation}
and when partials with respect to $z_k$ are taken on both sides of (\ref{E:sigg3}) analogous closed-form expressions are produced for higher order poles.  For example, a general fitting function of the form 
\begin{equation}\label{E:fitformHOP}
 \varphi = \sum\limits_{k=1}^{N_k}\ \sum\limits_{m=1}^{N_m(k)}\frac{{\nu}_{k,\,m}}{(z - z_k)^m}
\end{equation}
can be easily handled, where the ${\nu}_{k,\,m}$'s are the different pole strengths of various orders at distinct locations.  (Here ${N_m(k)}$ is the number of different types of poles at location $z_k$.)  Clearly when this form is substituted in place of (\ref{E:fitform}) into ${\Phi}\eq \|f - \varphi\|_{\sigma}^2$ and partials with respect to the pole parameters are taken, a set of closed-form linear equations results for these various higher-order pole strengths.

 In \cite{DIDACKSII}, point replicating expressions analogous to (\ref{E:rep1}) were given for both the interior and exterior of an $\mathbb{R}^3$ unit sphere, which in conjunction with the 
corresponding results for $\mathbb{R}^2$ units disks, leads to the assumption that analogous results can easily be derived for the interior and exterior of unit $\mathbb{R}^n$ hyper-spheres for $n >3$; however, it is unclear what the applications of such results might be so they are not considered here.

    With the formalism developed here, just as higher-order pole fits can be easily be done in the complex half-space setting, in the $\mathbb{R}^n$ half-space setting higher-order multipole fits can be done.  First, observe that when the partials of (\ref{E:fitform2}) with respect to the components of $\vec{X}$ are taken a fitting form results that is a combination of higher-order (electrostatic) multipole basis functions. Second, observe that partials of fundamental solutions with respect to the components of $\vec{X}$  and with respect to the components of the source coordinates $\vec{X}'_k$ differ by only a sign.  For example, in $\mathbb{R}^3$
\begin{equation}\label{E:grads}
\frac{\partial\ \,\,}{\partial\,x_k}\,\frac1{|\vec{X} - {\vec{X}}_k|} = - \,\frac{\partial\ }{\partial\,x}\,\frac1{|\vec{X} - {\vec{X}}_k|}\ . 
\end{equation}
Third, observe that when $\vec{P}_k = \vec{P}_k({\vec{X}}'_k)$ is assumed and the partials of both sides of (\ref{E:Bh}) are taken with respect to the components of source coordinates, then a closed form expression results on the RHS and these source partials can be taken inside the inner product on the left hand side (LHS).  In conjunction with the two other observations, this means that if a DIDACKS point replicating inner product exists for fundamental solutions, then closed-form inner-product expressions also exist for all higher-order multipole basis functions, such as dipoles and quadrupoles, and that interpolation fits for these higher order multipoles can easily be done.  (For further details see \cite{DIDACKSI} and \cite{DIDACKSII}.)  Due to fairly obvious connections between inverse-multiquadric radial basis function fits and higher-order multipole fits, the $\mathbb{R}^n$ half-space point replication conditions developed in Section~\ref{S:DHn} have clear implications for the reinterpretation of inverse-quadric radial basis function fits.

  Since the obvious connections between inverse-multiquadric radial basis functions and higher-order multipoles are invariably ignored (or have generally gone unrecognized), it is worth explicitly pointing them out here since they also imply that the associated radial basis functions fits can also be considered DIDACKS fits as well and there are advantages to doing this.  These connections are discussed in the next section.  In Section~\ref{S:RnSurf}, it is pointed out that a development similar to the DIDACKS half-space one can be carried out for the half-space surface norm and this implies that a corresponding inverse-multiquadric radial-basis function reinterpretation can be carried out for this norm as well.  Thus, from (\ref{E:Pcoll2}) and (\ref{E:Pcoll3}), this also implies that there are associated surface integral minimization conditions that hold implicitly for certain radial basis function fits.
 
\section{Reinterpretation of Radial Basis Function Fits}\label{S:RBF}

  This section gives a sketchy overview of techniques for reinterpreting radial basis function fits as DIDACKS fits.  The same issue is briefly addressed from a different perspective for a different norm setting in the next section. 

Consider a typical radial basis function fit to prescribed function values $f_k \eq f({\vec{Q}}_k)$ for $k = 1,\,2,\,3,\,\ldots,\,N_k\,,$ where ${\vec{Q}}_k \in \mathbb{R}^n$.  As normally performed, this $\mathbb{R}^n$ fit can be boiled down to the following three steps \cite{Buhmann,SandW}:
\begin{itemize}
\item
  First, an appropriate radial basis function shaping function $\phi(|\vec{X}|)$ is chosen, where $\vec{X} \in \mathbb{R}^n$. 
\item
  Second, it is assumed that the candidate fitting form is a linear combination of shifted shaping functions, so that
  \begin{equation}\label{E:fitformRBF}
 \varphi(\vec{X}) = \sum\limits_{k=1}^{N_k}\ {\mu}_{k}\,\phi(|\vec{X} - {\vec{Q}}_k|)\,\,,
\end{equation}
where $\mu_k \in \mathbb{R}$.
\item
 Third, and finally, the coefficients $\mu_k$ are determined directly by solving the point matching condition that corresponds to (\ref{E:Pcoll}):
\begin{equation}\label{E:RBFequation}
 \varphi({\vec{Q}}_k) = f_k\ \ \text{for}\ \ k = 1,\,2,\,3,\ldots,\,N_k\ .
\end{equation}
\end{itemize}   
One commonly used type of shaping functions are the inverse multiquadrics, which have the form \{see, for example, \cite{Buhmann} or equation (5.3) of \cite{SandW}\}:
\begin{equation}\label{E:multiquadrics}
\phi(r) = \frac1{(r^2 + a^2)^{\beta}}
\end{equation}
where $r = |\vec{X}|$, $\beta > n/2$ and $a$ is called the shape parameter.   To indicate how these inverse-multiquadric fits can be reinterpreted as DIDACKS fits, several specific examples
will be considered---from these examples the general strategies to be used in this reinterpretation process should be obvious.
 
  Before proceeding, notice that for both DIDACKS fits and radial basis function fits, that if individual basis functions are defined by $b\mathcal{B}_k$ or $b\phi$ for some constant $b$, rather than $\mathcal{B}_k$ or $\phi$, then the same final overall fitting expression results when a fit is carried out.  The constant $b$ will be called a basis function scaling factor.

  The value $\beta = n/2$ is also occasionally used in the form (\ref{E:multiquadrics}) to define the shaping function of choice.  When this particular shaping function is used to perform $\mathbb{R}^2$ radial basis function fits, (\ref{E:fitformRBF}) becomes
 \begin{equation}\label{E:fitformRBF2}
 \varphi\ls_{1/2}(\vec{X}) \eq \sum\limits_{k=1}^{N_k}\ \frac{{\mu}_{k}}{\sqrt{|\vec{X} - {\vec{Q}}_k|^2 + a^2}}\,\,,
\end{equation}
which corresponds to an $\mathbb{R}^3$ half-space point source fit of the form
\begin{equation}\label{E:beta12}
   \varphi(x,\,y,\,z) =
 \sum\limits_{k=1}^{N_k}\ \frac1{\sqrt{(x - x'_k)^2 + (y - y'_k)^2 + (z - z'_k)^2}}
\end{equation}
 provided $z'_k =$ constant, $\vec{Q}_k \eq (x'_k,\,y'_k)$, $z \eq |z'_k|$, $a \eq 2|z'_k|$ and $\varphi_{1/2}(x,\,y) \eq  \varphi(x,\,y,\,a/2)$.  [Notice that here, as well as in the sequel, this reidentification process really only requires that $z =$ constant in (\ref{E:beta12}).]

  Next observe that a dipole oriented along the $z-$axis in $\mathbb{R}^3$ yields a basis function of the form
 \begin{equation}\label{E:R3DP}
   \mathcal{B}_k(x,\,y,\,z) = \frac{\partial\ }{\partial z}
 \frac1{\sqrt{(x - x'_k)^2 + (y -y'_k)^2 + (z -z'_k)^2}} = -\frac{z -z'_k}{[(x - x'_k)^2 + (y -y'_k)^2 + (z -z'_k)^2]^{3/2}} 
\end{equation}
and thus if a basis function scaling factor of $1/(2z'_k)$ is introduced, the same reidentification process that was used to match up (\ref{E:beta12}) with (\ref{E:fitformRBF2}) can again be carried out for $\beta = 3/2$, when $n = 2$ in (\ref{E:multiquadrics}).  Taking higher-order partials results in like correspondences for $\beta = 5/2,\,7/2,/9/2$ etc. In all of these cases, observe that from (\ref{E:rep1}), (\ref{E:Pcoll2}) and (\ref{E:Pcoll3}), one can infer that the associated radial basis function fits also satisfy analogous energy minimization conditions---although the exact interpretation and forms of these corresponding energy minimization conditions is left to the reader.

   Next considered the generalization of this reinterpretation technique from $\mathbb{R}^2$ radial basis function fits to $\mathbb{R}^n$ radial basis function fits.  As a first step, for $\beta > 1/2$, notice that another obvious that way to obtain a correspondence between $\mathbb{R}^2$ radial basis functions and fundamental solutions exists for half-space fundamental solution fits in $\mathbb{R}^m$, for $m > 3$.  Thus consider the $\mathbb{R}^m$ form
\begin{equation}\label{E:funform}
 F = \frac{1\ \ }{|\vec{X} - \vec{X}'_k|^{m-2}}\,\,,
\end{equation}
which was introduced in (\ref{E:fitform2}).  Let $\vec{X}'_k = (x'_k,\,y'_k,\,0,\,0,\,\cdots,\,0,\,z'_k)^T$, then using the same reidentification technique as before it is clear that for $\varphi = \varphi(x,\,y,\,0,\,0,\,\cdots,\,0,\,|z'_k|)^T$, the associated $F$ corresponds to an $\mathbb{R}^2$ inverse multiquadric with $\beta = (m/2) - 1$.  This same reidentification procedure can also be carried out for higher dimensional radial basis function fits in an obvious way and thus corresponds to one of the desired general reinterpretation techniques.  Also, as before various partials of $F$ can be taken, which yields another reinterpretation procedure for $n > 2$.  As before, the interpretation of the associated energy minimization principles is left to the reader.

 There is one significant difference between the typical reidentification process epitomized by (\ref{E:beta12}) and (\ref{E:R3DP}):  Namely, when a partial derivative is taken, say with respect to $z$ as in (\ref{E:R3DP}), the resulting DIDACKS fit corresponds to matching point data for $\partial f/\partial z$ and not $f$ itself.  This, in turn, means that the associated inverse-multiquadric radial basis fit must be considered a DIDACKS fit that minimizes the difference of the appropriate multipole form and corresponding \emph{integrals} of $f$, which is a significant conceptual issue in itself.  However, there is a fairly direct way of injecting the fitting form (\ref{E:R3DP}) into surface integral dual-access collocation-kernel fits, as briefly mentioned in the next section.

\section{$\mathbb{R}^n$ Half-space Surface Integral Theory}\label{S:RnSurf} 
  
  For the complex setting, (\ref{E:sigg1}) implies that a standard (i.e., surface) integral norm point replication theory exists for $\mathbb{C}$ disks, while (\ref{E:siggRep}) implies that a similar development holds for $\mathbb{C}$ half-space.  Furthermore, from the complex to real norm mapping given in \cite{DIDACKSI}, it is easy to obtain corresponding $\mathbb{R}^2$ half-space relationships.  Section~\ref{S:SIDACKS} develops an analogous theory for $\mathbb{R}^n$ half-space (for $n > 2$).  In the complex setting the associated functional space setting was labeled a standard integral dual-access collocation-kernel space (SIDACKS).  For the $\mathbb{R}^n$ half-space setting the same acronym will be reused to mean a surface integral dual-access collocation-kernel space.  It is perhaps surprising that such a theory exists, but it is even more surprising that it has been (completely?) overlooked in the literature given its direct and obvious link to Poisson's integral formula for $\mathbb{R}^n$ half-space.  The resulting surface inner products between dipoles that are oriented along the upward pointing unit norm direction, such as (\ref{E:R3DP}), and some given $\mathbb{R}^n$ harmonic function $f$ yield a result proportional to a point evaluation of $f$ itself and not the partial of $f$, which implies an alternative reinterpretation of inverse-multiquadric radial basis function fits.  Moreover, this reinterpretation obviously implies that corresponding surface integral norm minimization conditions hold, in complete analogy to the energy-norm case.  Again this actual reinterpretation of inverse multiquadric radial basis function fits as $\mathbb{R}^n$ SIDACKS fits and the associated minimum norm implications are left to the reader to sort out. 

  There is one other significant problem area where SIDACKS theory is of special interest.  This is the problem area of downward continuation in $\mathbb{R}^n$ half-space.  There seems to be no existing appropriate physically motivated mathematical framework for rigorously handling this type of problem (although various mathematical formalisms do exist).  The general issue of using a (weighted) energy-norm approach was raised in Appendix~B of \cite{DIDACKSII} and it will be followed up in a subsequent article in this series.  Here downward continuation will be considered from a different perspective for half-space geometries in $\mathbb{R}^n$ for $n > 2$, although it is usually only the $\mathbb{R}^3$ half-space case that is of interest.  In general, the $\mathbb{R}^n$ half-space downward continuation problem is that of reconstructing the value of $f$ on some surface closer to $\partial\Hn$ from specified values at a surface that is further away from $\partial\Hn$ and labeled $\Sigma$.  Here it will be assumed that the surfaces in question are parallel to $\partial\Hn$.  Furthermore, for convenience, the surface where $f$ is to be reconstructed will be taken to be coincident with $\partial\Hn$ itself.

  Let $\varphi$ denote the full reconstruction that is to match $f$ on the surface $\Sigma$.
  First, observe that since the extremes of a harmonic function occur on the boundary, what one would ideally like to minimize is:
\begin{align}\label{E:SupNorm}
  &{\Phi}_{\infty} \eq \|\varphi - f\|{\ls}^2_{\infty}\ \ \text{where}\\
  &\|f\|{\ls}^2_{\infty}\eq \text{Sup}\ |f(\vec{P})|^2 \ \text{for all}\ \vec{P} \in \partial\Hn.
\end{align}
Notice here that $\|f\|^2{\ls}_{\infty}$ is the square of $\|f\|{\ls}_{\infty}\eq \text{Sup}\ |f(\vec{P})|$ for all $\vec{P} \in \partial\Hn$ and that it is somewhat hard to deal with computationally, so it is expedient to seek out an alternative minimization form.  Reasoning along similar lines, it would seem that the second best form to minimize would be 
\begin{align}\label{E:SupNorm2}
  &{\Phi}_{\sigma/\Hn} \eq \|\varphi - f\|{\ls}^2_{\sigma/\Hn}\ \ \text{where}\\
  &(f,\,g){\ls}_{\sigma/\Hn}\eq \int\limits_{\partial\Hn}\,\,f\,g\,\,d\,X\label{E:SupN}
\end{align}
and where $d\,X$ is defined in an obvious way in Section~\ref{S:DHn}.  Realistically, in general it is necessary to reconstruct $f$ from data values that are specified over only part of $\Sigma$ and where noise is present in the data that is available.  The issue of noisy data is ignored here, but if appropriate inner products can be evaluated from sufficiently accurate available data, then $\varphi$ can be used to do downward continuation.  In Section~\ref{S:SIDACKS}, it is shown that two types of surface inner products can be theoretically evaluated.  One where the inner product involves dipoles and is naturally associated with point evaluation on $\Sigma$ and one where the inner product involves fundamental solutions, but that requires information about $f$ that is in the region above $\Sigma$ [in particular, it requires line integrals of $f$ outward along the (upward pointing) normal direction of $\partial\Hn$ from some specified point of $\Sigma$].

\section{Point Replication Conditions for Complex Half-space}\label{S:Recap} 

  In this section the replication conditions for half-space corresponding to (\ref{E:sigg3}), (\ref{E:CollPropD}), (\ref{E:psiRep}) and (\ref{E:xi}) are derived.  There are three general strategies for obtaining these corresponding relationships:
\begin{enumerate}
\item[(a)]
  Take the appropriate limits of (\ref{E:sigg3}), (\ref{E:CollPropD}), (\ref{E:psiRep}) and (\ref{E:xi}) by considering what happens when $z_k \approx (0,\,i)$ and then translating to new coordinates.
\item[(b)]
  Develop entirely new relationships using standard analysis techniques.
 \item[(c)] 
 Recast existing RKHS complex half-space (or half-space boundary) kernels into dual-access collocation-kernel form and then derive the corresponding point replication conditions.
\end{enumerate}
  The last approach was addressed in \cite{DIDACKSI} where, for example, various specific existing kernels were analyzed for both the interior and exterior of complex unit circles and unit disks; moreover, it was also pointed out there that a general correspondence exists between a Bergman kernel for any specified complex domain and a complex Dirichlet integral reproducing kernel defined over the same domain, so that Bergman kernel results for half-space can be recast as Dirichlet integral dual-access collocation-kernel results.  Nevertheless, unless the study of Bergman kernels is the primary focus, approach (c) is not really as enlightening as either approach (a) or (b), so it will not be followed up here.  (As an aside, it is an unstated implicit goal of the DIDACKS programme to develop all the associated kernels independently and then recast them in RKHS form in order to verify that no significant RKHS possibilities have been overlooked.)  This leaves a choice between approach (a) and (b).  Although both approaches are instructive and there are some subtleties involved in carrying out either one; the second approach is ultimately the more satisfying and convincing one so it is the main one opted for here.

  Clearly the unmodified series form (\ref{E:series1}) is inappropriate for the half-space case due to its behavior at the origin of $\Hc$.  Thus for both the standard integral complex half-space setting and the Dirichlet integral complex half-space setting, it is assumed that all admissible analytic functions $f$ have a power series representation of the form
\begin{equation}\label{E:seriesHc}
 f(z) = \sum\limits_{j=1}^{\infty} \frac{a_j}{(z - d)^{j}}
\end{equation}
in $\Hc$ for bounded $d \in \hc$ and $f$ that tails off ``sufficiently fast'' as $|z| \rightarrow \infty$.  (Here ``sufficiently fast'' should be obvious in the sequel from the context.)

    Looking ahead, for each point replication condition it is necessary to consider, not only the definition of the inner product, but of the fundamental reproducing kernel, $F_k$, as well.  The point is that although it may be assumed, for example, that $F_k$ has the form of a simple pole, for convenience it may be necessary to include an overall basis function scaling factor so that for simple poles the general kernel form assumed will be $F_k = b/(z - z_k)$ for $z_k \in \hc$.  

\pagebreak
\begin{center} 
\textbf{\underline{Standard integral norm for the complex half-space setting}}
 \hfil\\
 \hfil\\
\end{center}

  For $z \in \Hc$ the natural analog of (\ref{E:sigg1}) to consider is
\begin{equation}\label{E:siggH1}
(g,\,f){\ls}_{\sigma/2} \eq  \frac1{2\pi} \int\limits_{x=-\infty}^{\infty} [(g(z))^*\,f(z)]\Big|_{y=0} d\,x\ . 
\end{equation}
Here the integral $\|f\|^2_{\sigma/2}$ must be bounded.  Also observe that from the assumed series form (\ref{E:seriesHc}) it follows that
\begin{equation}\label{E:LineInt}
\lim\limits_{R\rightarrow\infty}\int\limits_{\theta=0}^{\pi} [(g(z))^*\,f(z)]\Big|_{r=R} d\,\theta\ = 0\ .
\end{equation}

  Next consider the Cauchy integral formula 
\begin{equation}\label{E:Cauchy}
f(\zeta) = \frac1{2\pi i}\oint\limits_{\gamma}\frac{f(z)}{z - \zeta}\,\,d\,z\ ,
\end{equation}
where $f$ is analytic in the simply connected region bounded by the closed curve $\gamma$ and $\zeta$ is in this same region.  Taking (\ref{E:LineInt}) into account and choosing $\gamma$ in (\ref{E:Cauchy}) to be the path from $-R$ to $R$ along the $x-$axis and then from  $\theta = 0$ to $\theta = \pi$ along a circular arc of radius $R$, immediately gives 
\begin{equation}\label{E:Cauchy2}
f(\zeta) = \frac1{2\pi i}\int\limits_{x=-\infty}^{\infty}\Blbrac\frac{f(z)}{z - \zeta}\Brbrac\Bigg|_{y=0}\,\,d\,x
\end{equation}
for admissible functions $f$ when $R \rightarrow \infty$.  The exact choice of the form for $F_k$ remains, but a comparison of (\ref{E:siggH1}) and (\ref{E:Cauchy2}) is suggestive.

   In \cite{DIDACKSI} it was pointed out that $1/(z - z_k)$ corresponds to an $R^2$ dipole term oriented along the $x-$axis and that $i/(z - z_k)$ corresponds to an $R^2$ dipole term oriented along the $y-$axis.  From the orientation chosen for the half-space $\Hc$ itself (i.e., $y>0$), it is obvious that fundamental kernels should correspond to $R^2$ dipoles oriented along the $y-$axis.  Thus introducing a basis function scaling factor of $b = i$ results in the choice
\begin{equation}\label{E:Fkz}
 F_k(z) \eq \frac{i}{z - z_k}\ .
\end{equation}
Substitution of this expression for $g$ into (\ref{E:siggH1}) immediately yields
\begin{equation}\label{E:siggH2}
(\,i(z - z_k)^{-1},\,f){\ls}_{\sigma/2} =  \frac{-i\,\,}{2\pi}\!\!\int\limits_{x=-\infty}^{\ \ \infty} \frac{f(x,\,0)}{(x - z_k^*)}\,\, d\,x\ . 
\end{equation}
(As an aside, since the term ``fundamental solution'' is not normally used in the complex setting, the term ``fundamental kernel'' is used here to mean a simple pole; alternatively, in the $\mathbb{R}^2$ setting logarithmic sources are fundamental solutions so from the natural correspondence implied by standard harmonic completion \cite{DIDACKSI} one could take the term ``fundamental solution'' in $\mathbb{C}$ to be a logarithmic term, but $1(z - z_k)$ is much more ubiquitous than logarithmic terms are in $\mathbb{C}$ so a new phrase with a new meaning is introduced here to simply by-pass this issue.)
Introducing the half-space relationship $q_k \eq z_k^*$ and comparing (\ref{E:siggH2}) with (\ref{E:Cauchy2}) gives the desired result:
\begin{equation}\label{E:siggRep}
(F_k,\,f){\ls}_{\sigma/2} = f(q_k) = f(z^{*}_k)\ .
\end{equation}
Here $q_k$ is the half-space analog of the unit disk interior and exterior conjugate involution point of $z_k$: $p^*_k = 1/z^*_k$, which should be obvious from a consideration of the limiting process mentioned at the first of this section in connection with approach (a). 

\pagebreak
\begin{center} 
 \hfil\\
\textbf{\underline{Dirichlet integral norm for the complex half-space setting}}
 \hfil\\
\end{center}

  The natural half-space analog of (\ref{E:CplxDirInt}) to consider is
\begin{equation}\label{E:siggH3}
(g,\,f){\ls}_{D/2} \eq  \frac1{2\pi} \int\limits_{x=-\infty\ }^{\ \ \infty}\int\limits_{y=0}^{\ \ \infty} \bigg(\frac{d\,g}{d\,z}\bigg)^{*}\bigg(\frac{d\,f}{d\,z}\bigg)\,  d\,y\ d\,x\ . 
\end{equation}

  First, consider the choice of the overall basis function scaling factor.  In (\ref{E:Fkz}) the choice $b = i$ was justified on the grounds that it corresponds to a $\mathbb{R}^2$ dipole oriented along the $y-$axis.  This choice could have been argued in perhaps a more precise way by looking at the limiting process of ($\ref{E:sigg3}$) mentioned at the first of this section in connection with an alternative derivation process [approach (a)].  In this limit for ($\ref{E:sigg3}$) $p_k^* \approx i$, so it is obvious that an extra factor of $i$ enters in the final limit from the RHS of (\ref{E:sigg3}), which corresponds to a choice of $b = i$.  When the same limiting process of (\ref{E:CollPropD}) is considered, then $(p_k^*)^2 \approx -1$ occurs in the limit of the RHS of (\ref{E:CollPropD}) and thus, aside from a possible overall sign difference that can be accounted for later, there is no need to introduce a basis function scaling factor.  Hence for convenience the choice
\begin{equation}\label{E:Fk}
 F_k(z) \eq \frac1{(z - z_k)}
\end{equation}
is made here and the inner product of interest becomes
\begin{equation}\label{E:DD2}
   (F_k,\,f){\ls}_{D/2} \eq  \frac{-1}{2\pi} \int\limits_{x=-\infty\ }^{\ \ \infty}\int\limits_{y=0}^{\ \ \infty} \frac{f'(z)}{(x - iy - q_k)^2}\,  d\,y\ d\,x\ , 
\end{equation}
where, as before, $q_k = z_k^*$ for the half-space setting here [see the comment after (\ref{E:siggRep})].

  Observe that due to the factor of $- iy$ rather than $iy$ occuring in the denominator of the integrand on the RHS of (\ref{E:DD2}), the Cauchy integral formula cannot be applied directly, and, furthermore, it is hard to transform this factor of $- iy$ away while still preserving the analytic character of the rest of this integrand; hence, a certain amount of formal manipulation seems to be needed.  Thus, first let
\begin{equation}\label{E:Kx}
 \mathcal{K}(x) \eq \int\limits_{y=0}^{\ \ \infty} \frac{f'(z)}{(x - iy - q_k)^2}\,\,d\,y
\end{equation}
so that (\ref{E:DD2}) becomes
 \begin{equation}\label{E:KD}
   (F_k,\,f){\ls}_{D/2} =  \frac{-1\,\,}{2\pi}\!\! \int\limits_{x=-\infty\ }^{\ \ \infty}\mathcal{K}(x)\,\,d\,x\ .
\end{equation}
Since 
\begin{equation*}\notag
 \frac{\partial\ }{\partial y} \frac{f'}{(x - iy - q_k)} = \frac{\frac{\partial f'}{\partial y}}{(x - iy - q_k)} + \frac{i\,f'}{(x - iy - q_k)^2}
\end{equation*}
and from the Cauchy-Riemann equations
\begin{equation*}\notag
 \frac{\partial f'}{\partial y\,\,} = i\frac{\partial f'}{\partial x\,\,}
\end{equation*}
(i.e., if $g = u + iv$ then $[\partial u/\partial y] + i[\partial v/\partial y] = -[\partial v/\partial x]+ i[\partial u/\partial x]$) it follows that
\begin{align}\label{E:K1}
\mathcal{K}(x) &= -i \int\limits_{y=0}^{\ \ \infty} \frac{\partial\ }{\partial y}\Blbrac\frac{f'(z)}{(x - iy - q_k)}\Brbrac\,\,d\,y + i \int\limits_{y=0}^{\ \ \infty} \frac{\frac{\partial f'}{\partial y\,\,}}{(x - iy - q_k)}\ d\,y\notag \\
&= \frac{if'(x,\,0)}{(x - q_k)} -  \int\limits_{y=0}^{\ \ \infty} \frac{\frac{\partial f'}{\partial x\,\,}}{(x - iy - q_k)}\ d\,y\ .
\end{align}

  Likewise
\begin{equation*}\notag
 \frac{\partial\ }{\partial x} \frac{f'}{(x - iy - q_k)} = \frac{\frac{\partial f'}{\partial x\,\,}}{(x - iy - q_k)} - \frac{f'}{(x - iy - q_k)^2}
\end{equation*}
so that
\begin{equation}\label{E:K2}
\mathcal{K}(x) = - \int\limits_{y=0}^{\ \ \infty} \frac{\partial\ }{\partial x}\Blbrac\frac{f'(z)}{(x - iy - q_k)}\Brbrac\,\,d\,y + \int\limits_{y=0}^{\ \ \infty} \frac{\frac{\partial f'}{\partial x\,\,}}{(x - iy - q_k)}\ d\,y\ .
\end{equation}
Adding (\ref{E:K2}) to the extreme LHS and RHS of (\ref{E:K1}) and dividing by two yields:
\begin{equation*}\notag
\mathcal{K}(x) = \frac{if'(x,\,0)}{2(x - q_k)} - \frac12\int\limits_{y=0}^{\ \ \infty} \frac{\partial\ }{\partial x}\Blbrac\frac{f'(z)}{(x - iy - q_k)}\Brbrac\,\,d\,y  .
\end{equation*}
Thus after the indicated evaluations of the resulting double integral are performed the following desired expression results
\begin{equation}\label{E:K3}
\int\limits_{x=-\infty\ }^{\ \ \infty}\mathcal{K}(x)\,d\,x = \frac{i}2\int\limits_{x=-\infty\ }^{\ \ \infty}\frac{f'(x,\,0)}{(x - q_k)}\,\,d\,x = -\frac{\pi}{2\pi i}\int\limits_{x=-\infty}^{\infty}\Blbrac\frac{f'(z)}{z - q_k}\Brbrac\Bigg|_{y=0}\,\,d\,x = -\pi f'(q_k)\,\,,
\end{equation}
where the Cauchy integral formula given by (\ref{E:Cauchy2}) was used the evaluate the RHS here.

    Substituting this result into (\ref{E:KD}) immediately gives the end result:
\begin{equation}\label{E:KD3}
   (\,(z - z_k)^{-1},\,f){\ls}_{D/2} =  \frac{1}{2}\,f'(q_k) = \frac12\frac{d\,f}{d\,z}\Bigg|_{z=z^*_k}.
\end{equation}

\begin{center} 
 \hfil\\
\textbf{\underline{Complex half-space logarithmic Dirichlet-integral inner-product expressions}}
 \hfil\\
\end{center}

  The inner product is the same $\Hc$ one just analyzed; furthermore, $z_k$ and $q_k$ have the same meaning so that $q_k = z^*_k$ with ${\text{Im}}\,\{q_k\} > 0$.  Now, however, the following basis function will be considered, which is the obvious analog of (\ref{E:psilog}):
\begin{equation}\label{E:psilog2}
  \psi_k(z) \eq \ln\, \frac{z + i}{(z - z_k)}\ .
\end{equation}
The following analog of (\ref{E:xibasis1}) will also be considered:
\begin{equation}\label{E:xibasis2}
 {\xi}_k(z) \eq \ln\,\frac1{z - z_k} - \ln\,\frac1{z - z'_k}\, = \,\ln\,\left(\frac {z - z'_k}{z - z_k}\right)\ ,
\end{equation}
which is a generalization of (\ref{E:psilog2}).  Here  ${\text{Im}}\,\{z'_k\} < 0$ and it is useful to introduce $q'_k \eq (z'_k)^{*}$.  Although the factor of $i$ occuring on the RHS of (\ref{E:psilog2}) looks strange and it occurs because $\psi_k(z)$ is not translation invariant [while 
${\xi}_k(z)$ is], it is expedient to derive expressions for it first.

  Observe that
\begin{equation}\label{E:psilog3}
  \frac{d\,\psi_k}{d\,z\ } =  \frac1{(z + i)} -  \frac1{(z - z_k)}\ .
\end{equation}
Thus from (\ref{E:siggH3}), the inner product expression of interest is
\begin{equation}\label{E:psiRep2}
  (\psi_k,\,f){\ls}_{D/2} = \frac{1}{2\pi} \int\limits_{x=-\infty\ }^{\ \ \infty}\int\limits_{y=0}^{\ \ \infty} \left(\frac1{x - iy - i} - \frac1{x - iy - q_k}\right)\,{f'(z)}\,\,\, d\,y\ d\,x\ . 
\end{equation}

   For the sake of variety, consider a Wirtinger calculus based derivation approach \cite{DIDACKSI}.  First, since the Wirtinger calculus is based on the concept of conjugate variables, recall that the notation $(z,\,\bar{z})$ was reserved for this pair of variables in \cite{DIDACKSI} so that, in general, $f = f(z,\,\bar{z})$.   Since $f$ is analytic it does not depend on $\bar{z}$ and 
\begin{equation}\label{E:Wirt}
 \frac{d\,f}{d\,z} = \frac{\partial\,f}{\partial\,z}
\end{equation}
so that the usual Wirtinger derivatives \{c.f., equation (1) of \cite{DIDACKSI}\}
\begin{equation*}\notag
 \frac{\partial\ }{\partial\, z} = \frac12\left(\frac{\partial\,\ }{\partial\,x} - i\frac{\partial\,\ }{\partial\,y}\right)
\end{equation*}
 can be used to reexpress (\ref{E:psiRep2}):
\begin{equation}\label{E:psiRep3}
  (\psi_k,\,f){\ls}_{D/2} = \frac{1}{2\pi} \int\limits_{x=-\infty\ }^{\ \ \infty}\int\limits_{y=0}^{\ \ \infty} \left(\frac1{x - iy - i} - \frac1{x - iy - q_k}\right)\, \frac12\left(\frac{\partial\,f}{\partial\,x} - i\frac{\partial\,f}{\partial\,y}\right)
\,\,\, d\,y\ d\,x\ . 
\end{equation}
Now for any constant $a \in \mathbb{C}$
\begin{equation*}\notag
 \ \ \frac{\partial\,\ }{\partial\,x}\,\frac{f}{(x - iy - a)} =   \frac{-f}{(x - iy - a)^2} + \frac1{(x - iy - a)}\frac{\partial\,f}{\partial\,x}
\end{equation*}
and 
\begin{equation*}\notag
 -i\frac{\partial\,\ }{\partial\,y}\,\frac{f}{(x - iy - a)} =   \frac{f}{(x - iy - a)^2} - i \frac1{(x - iy - a)}\frac{\partial\,f}{\partial\,y}
\end{equation*}
 so 
\begin{equation}\label{E:psiRep4}
  (\psi_k,\,f){\ls}_{D/2} = \frac{1}{2\pi} \int\limits_{x=-\infty\ }^{\ \ \infty}\int\limits_{y=0}^{\ \ \infty} \frac12\left(\frac{\partial\,\ }{\partial\,x} - i\frac{\partial\,\ }{\partial\,y}\right)\left(\frac{f}{x - iy - i} - \frac{f}{x - iy - q_k}\right)\, 
\,\,\, d\,y\ d\,x\ . 
\end{equation}
Obviously 
\begin{equation*}\notag
 \int\limits_{x=-\infty\ }^{\ \ \infty}\left(\frac{\partial\,\ }{\partial\,x}\right)\left(\frac{f}{x - iy - i} - \frac{f}{x - iy - q_k}\right)\,\, d\,x = 0 
\end{equation*}
and 
\begin{equation*}\notag
\int\limits_{y=0}^{\ \ \infty} \left(i\frac{\partial\,\ }{\partial\,y}\right)\left(\frac{f}{x - iy - i} - \frac{f}{x - iy - q_k}\right)\, 
\,\,\, d\,y\  = - \left(\frac{i}{x - i} - \frac{i}{x - q_k}\right)f(x,\,0) 
\end{equation*}
so 
\begin{equation}\label{E:psiRep5}
  (\psi_k,\,f){\ls}_{D/2} = \frac{1}{2\pi} \int\limits_{x=-\infty\ }^{\ \ \infty} \frac{i}2\left(\frac{f(x,\,0)}{x - i} - \frac{f(x,\,0)}{x - q_k}\right)\,\, d\,x\ , 
\end{equation}
which involves the same form of integrals that occured on the RHS of (\ref{E:siggH2}) or (\ref{E:Cauchy2}).  Hence, from (\ref{E:Cauchy2})
\begin{equation}\label{E:psiRep6}
  (\psi_k,\,f){\ls}_{D/2} =  \frac12\,f(q_k) - \frac12\,f(i)\ .
\end{equation}
Retracing the steps in the above derivation it is obvious that
\begin{equation}\label{E:xi2}
({\xi}_k,\,f){\ls}_{D/2} = \frac12\,f(q_k) - \frac12\,f(q'_k)\ ,
\end{equation}
in accord with the complex half-space limit of (\ref{E:xi}).

\section{Dirichlet Integral Replication Expressions for $\mathbb{R}^n$ Half-space}\label{S:DHn}

   This section derives $\mathbb{R}^n$ half-space results analogous to (\ref{E:rep1}).  As previously indicated, the $\mathbb{R}^2$ half-space inner-product expressions can be obtained in a fairly straightforward manner from the Dirichlet integral complex half-space results, so only the case $n > 2$ will be considered.

   First, consider the question of overall notational conventions.  While there are several older reference works that describe general $\mathbb{R}^n$ potential theory, such as \cite{Richards}, it is useful to stick closely to more modern treatments since several notational issues arise in $\mathbb{R}^n$ that do not arise in  $\mathbb{R}^3$ and it is a good idea to have a  reference that is readily available.  Since \cite{AxlerEtAll} gives a fairly broad treatment of $\mathbb{R}^n$ potential theory at the introductory level, since it contains all of the necessary results, since it is in print and widely available and, finally, since it is well written it will be used as the primary reference here; moreover, to circumvent the issues just noted, the notation adopted here will be the same as that used in \cite{AxlerEtAll} except for a few indicated or obvious exceptions.
    
  Next, consider the task of reexpressing the Dirichlet integral given on the RHS of (\ref{E:DnIP}).  Three mathematical tools are useful in this endeavor:
\begin{itemize}
\item
 The divergence theorem \{\cite[p. 4]{AxlerEtAll}: Equation 1.2\}
  \begin{equation}\label{E:DivThm}
   \int\limits_\Omega \nabla\cdot\vec{W}\,\,d\,V =  \int\limits_{\partial\Omega} \hat{n}\cdot\vec{W}\,\,d\,S
   \end{equation}
where $\vec{W}$ is a vector field in $\mathbb{R}^n$ [$\vec{W} = \vec{W}(\vec{X})$ for $\vec{W}$ and $\vec{X} \in \mathbb{R}^n$] and $\hat{n} = \hat{n}(\vec{X})$ is the outward pointing unit normal vector associated with the boundary surface ${\partial\Omega}$, which has a differential hyper-surface area of $d\,S$.
\item
 The product rule for the Laplacian given by \{\cite[p. 13]{AxlerEtAll}: Equation 1.19\}
  \begin{equation}\label{E:ProductRule}
  {\nabla}^2(uv) = u{\nabla}^2v + 2{\nabla}u\cdot{\nabla}v + v{\nabla}^2u\ .
  \end{equation}
\item
  Green's identity, which can be written as \{\cite[p. 4]{AxlerEtAll}: Equation 1.1\}
 \begin{equation}\label{E:GreenThm}
    \int\limits_{\partial\Omega} \left(u\tfrac{\partial v}{\partial \hat{n}} - v\tfrac{\partial u}{\partial \hat{n}}\right)\,\,d\,S = \int\limits_\Omega (u{\nabla}^2v - v{\nabla}^2u)\,\,d\,V\ . 
   \end{equation}
\end{itemize}
If ${\nabla}^2\phi = {\nabla}^2\psi = 0$ holds over all of $\Omega$, then from (\ref{E:ProductRule})
\begin{equation*}\notag
  {\nabla}\phi\cdot{\nabla}\psi = \frac12{\nabla}^2(\phi\psi) = \frac12{\nabla}\cdot{\nabla}(\phi\psi)
\end{equation*}
and it follows from (\ref{E:DivThm}) that
\begin{equation}\label{E:DnInt2}
\text{D}[\phi,\,\psi,\,1,\,\Omega] =  \frac12\int\limits_\Omega {\nabla}\cdot{\nabla}(\phi\psi)\,\,d\,V  = \frac12\int\limits_{\partial\Omega} \hat{n}\cdot{\nabla}(\phi\psi)\,\,d\,S = \frac12\int\limits_{\partial\Omega}\frac{\partial\ }{\partial \hat{n}}(\phi\psi)\,\,d\,S\ .
\end{equation}
Further it follows from (\ref{E:GreenThm}) that 
\begin{equation}\label{E:GreenThm2}
    \int\limits_{\partial\Omega} \phi\tfrac{\partial \psi}{\partial \hat{n}}\,\,d\,S =  \int\limits_{\partial\Omega} \psi\tfrac{\partial \phi}{\partial \hat{n}}\,\,d\,S
   \end{equation}
so that finally
\begin{equation}\label{E:DnInt3}
\text{D}[\phi,\,\psi,\,1,\,\Omega] = \int\limits_{\partial\Omega}\psi\tfrac{\partial\phi}{\partial \hat{n}}\,\,d\,S\ .
\end{equation}

  Consider the specialization of (\ref{E:DnInt3}) to $\Hn$.  Following \cite{AxlerEtAll}, let $\vec{Z} \in \Hn \subset \mathbb{R}^n$ denote the vector variable of interest and reexpress it as a combination of two parts: For $n - 1 \geq j \geq 1$ let $x_j = z_j$ and for $j = n$ let $y = z_n$, so that $\vec{Z} = (z_1,\,z_2,\,z_3,\,\cdots,\,z_{n-1},\,z_n)^T = (x_1,\,x_2,\,x_3\cdots,\,x_{n-1},\,y)^T = [\,\vec{X}^T\,|\,y\,]^T$ where $\vec{X} \in \mathbb{R}^{n-1}$ and $\vec{X} \eq (x_1,\,x_2,\,x_3\cdots,\,x_{n-1})^T$.  \{Here $\vec{Z}$ was written as $[\,\vec{X}^T|\,\,y\,]^T$ rather than in the equivalent form $(\vec{X}^T,\,y)^T$ in order to emphasize the fact that a block matrix partition is implied.\}  Further, let $d\,X = d\,x_1\,d\,x_2\,d\,x_3\,\cdots\,d\,x_{n-1}$ and let
\begin{equation*}\notag
    \int\limits_{\mathbb{R}^{n-1}} = \int\limits_{\partial \Hn} = \int\limits_{x_1=-\infty\ }^{\ \ \infty}\int\limits_{x_2=-\infty\ }^{\ \ \infty} \int\limits_{x_3=-\infty\ }^{\ \ \infty}\ \mathbf{\cdots}\  \int\limits_{x_{n-1}=-\infty\ }^{\ \ \infty}\ .
\end{equation*}  
Thus, since $\partial /\partial \hat{n}= -\partial /\partial y$ for the half-space $\Hn$ given by $y \geq 0$, it follows from (\ref{E:DnInt3}) that
\begin{equation}\label{E:DnInt4}
\text{D}[\phi,\,\psi,\,1,\,\Hn] = -\int\limits_{\mathbb{R}^{n-1}}\psi\frac{\partial\phi}{\partial y}\,\,d\,X\ .
\end{equation}

\newcommand{\PH}{P_{\smallindexes{\!\text{H}}}}

   The strategy will be to compare the results from an evaluation of (\ref{E:DnInt4}) for a suitable choice of $\phi$ with Poisson's integral formula.  Temporally leaving aside the choice of $\phi$ to be made, consider Poisson's integral formula for $\mathbb{R}^n$ half-space.  From \cite[p. 9]{AxlerEtAll}:
\begin{equation}\label{E:Poissonint}
 \psi(\vec{z}) = \int\limits_{\mathbb{R}^{n-1}}\PH\,\psi(\vec{t},0)\,d\,t
\end{equation}
where
$\vec{t} \in \mathbb{R}^{n-1}$ is restricted to $\partial \Hn$, just as $\vec{X} \in \mathbb{R}^{n-1}$ was and $d\,t$ is defined in complete analogy to $d\,X$.  In (\ref{E:Poissonint}) $\PH = \PH(\vec{Z},\,\vec{t}\,\,)$ has the form \cite[p. 145]{AxlerEtAll}
\def\vt{\vec{t}\,\,}
\def\vtT{\vec{t}^{\,\,T}}
\begin{equation}\label{E:PH}
 \PH(\vec{Z},\,\vt) = \frac{ c_n\,y\ }{(|\vec{X} - \vt|^2 + y^2)^{n/2}}\,\,
\end{equation}
where \cite[p. 145]{AxlerEtAll}
\begin{equation}\label{E:Cn}
c_n = \frac2{nV(B)}
\end{equation}
with \cite[p. 245]{AxlerEtAll}
\begin{equation}\label{E:VB}
V(B) = \frac{{\pi}^{n/2}}{\Gamma(\tfrac{n}{2} + 1)} 
\end{equation}
so
\begin{equation}\label{E:Poissonint2}
 \psi(\vec{z}\,) = c_ny\int\limits_{\mathbb{R}^{n-1}}\frac{\psi(\vec{t},0)}{(|\vec{X} - \vt|^2 + y^2)^{n/2}}\,d\,t\ .
\end{equation}

  Next consider the choice of basis function and associated global basis scaling factor.  From the discussion surrounding (\ref{E:fitform2}) it is clear that the general form of interest is
\begin{equation}\label{E:fitform5}
 F(\vec{Z},\vec{S}\,) = \frac{d_n\ \ }{|\vec{Z} - \vec{S}\,|^{n-2}}
\end{equation}
for some appropriate basis scaling factor $d_n$, 
where $\vec{Z} \in \Hn$ and $\vec{S} \in \hn$.
For example, with this choice of fundamental solution basis function, (\ref{E:fitform2}) directly becomes
\begin{equation}\label{E:fitform6}
 \varphi = \sum\limits_{k=1}^{N_k}\ \frac{{\mu}_k\ \ }{|\vec{X} - \vec{X}'_k|^{n-2}} = \frac1{d_n}\sum\limits_{k=1}^{N_k}\ {\mu}_k\,F(\vec{X},\,\vec{X}'_k)
\end{equation}
in the notation used there.  Although \cite{AxlerEtAll} and \cite{Richards} assume $d_n < 0$, to simplify matters here it is useful to take $d_n > 0$.  The negative of the basis function scaling factor used in \cite[p. 193]{AxlerEtAll} leads to the choice
\begin{equation}\label{E:dScale}
 d_n = \frac1{(n - 2)nV(B)} = \frac{c_n}{2(n - 2)}\,\,,
\end{equation}
which is the one adopted here.

  In analogy to the block matrix partition of $\vec{Z}$ into $\vec{X}$ and $y$, let $\vec{S} = [\,{\vtT}|\,w\,]^T$ where $w \in \mathbb{R}$ is negative.  Then by a straightforward evaluation
\begin{equation}\label{E:fitform8}
 \frac{\partial\ }{\partial y}F(\vec{Z},\vec{S}\,) = -\frac{c_n}{2}\frac{(y - w)\ \,}{|\vec{Z} - \vec{S}\,|^{n}}\ .
\end{equation}
 Substituting $\phi = F(\vec{Z},\,\vec{S}\,)$ into (\ref{E:DnInt4}) thus yields
\begin{equation}\label{E:DnInt9}
\text{D}[\phi,\,\psi,\,1,\,\Hn]\, = \,\frac{c_n}{2}\int\limits_{\mathbb{R}^{n-1}}\Blbrac\frac{(y - w)\,\psi}{|\vec{Z} - \vec{S}\,|^{n}}\Brbrac\Bigg|_{y=0}\,d\,X\, = \,\frac{c_n}{2}|w|\int\limits_{\mathbb{R}^{n-1}}\frac{\psi(\vec{X},0)}{(|\vec{X} - \vec{t}|^2 + w^2)^{n/2}}\,\,d\,X\ .
\end{equation}
Introducing $\vec{P} \eq [\,\vtT\,|-w\,]^T$ so that $\vec{P} \in \Hn$ and comparing (\ref{E:DnInt9}) with (\ref{E:Poissonint2}) immediately gives (after an obvious change of variables)
\begin{equation}\label{E:DnRep}
\text{D}[F,\,\psi,\,1,\,\Hn] = \frac{1}{2}\,\psi(\vec{P}) .
\end{equation}

  For $n = 3$, $V(B) = 4\pi/3$ and $d_n = d_3 = 1/(4\pi)$ so that $F(\vec{X},\,\vec{S}\,) = 1/(4\pi|\vec{X} - \vec{S}\,|)$ and $1/\ell_k = 4\pi\,F(\vec{X},\,\vec{X}'_k)$.  Thus (\ref{E:DnRep}) gives 
\begin{equation}\label{E:DnRep3}
\text{D}[{\ell}_k^{-1},\,\psi,\,1,\,{\text{H}^3}]\, = \,{2\pi}\,\psi(\vec{P}_k)
\end{equation}
in accord with (\ref{E:rep1}).

\section{Surface Integral Replication Expressions for $\mathbb{R}^n$ Half-space}\label{S:SIDACKS}

  As discussed in Section~\ref{S:RnSurf} this section derives replication expressions for the 
 $\mathbb{R}^n$ half-space surface integral setting (i.e., SIDACKS).  Many of the results and much of the notation from the preceeding section are reused here.  Thus using the notation introduced prior to (\ref{E:DnInt3}), the inner product of interest here is
\begin{equation}\label{E:SI1}
(f,\,g){\ls}_{\sigma/\Hn} \eq \int\limits_{\mathbb{R}^{n-1}}\![f\,g]\Big|_{y=0}\,\,\,d\,X\,\,,
\end{equation}
in accord with (\ref{E:SupN}).  

  From (\ref{E:GreenThm2})
\begin{equation}\label{E:GreenThm8}
    \int\limits_{\mathbb{R}^{n-1}} f\tfrac{\partial g}{\partial y}\,\,d\,X =  \int\limits_{\mathbb{R}^{n-1}} g\tfrac{\partial f}{\partial y}\,\,d\,X
\end{equation}
and from (\ref{E:DnInt4})
\begin{equation}\label{E:SurfInt4}
(g,\,{\partial f}/{\partial y}){\ls}_{\sigma/\Hn} = -\text{D}[g,\,f,\,1,\,\Hn]\ .
\end{equation}
Which from (\ref{E:DnRep}) implies that
\begin{equation}\label{E:SurfInt5}
(F,\,{\partial f}/{\partial y}){\ls}_{\sigma/\Hn} = -\frac12f(\vec{P})\,\,,
\end{equation}
where $F$ is given by (\ref{E:fitform5}).  From (\ref{E:GreenThm8}) this can be rewritten as
\begin{equation}\label{E:SurfInt6}
(F,\,{\partial f}/{\partial y}){\ls}_{\sigma/\Hn} = ({\partial F}/{\partial y},\,f){\ls}_{\sigma/\Hn} = (F_y,\,f){\ls}_{\sigma/\Hn} = -\frac12f(\vec{P})\,\,,
\end{equation}
where 
\begin{equation}\label{E:SurfInt7}
F_y \eq \frac{\partial F}{\partial y}\ .
\end{equation}
The last two expressions in (\ref{E:SurfInt6}) can be used to perform a SIDACKS dipole fit directly and, as such, can be used to reinterpret inverse-quadric radial basis function fits as previously noted; moreover, these expressions can also clearly be used to perform downward continuation.

  In many cases it is also possible to do SIDACKS fundamental solution based inner product fits.  From (\ref{E:SurfInt6})
\begin{equation}\label{E:SurfInt8}
({\partial F}/{\partial y},\,f){\ls}_{\sigma/\Hn} = -({\partial F}/{\partial w},\,f){\ls}_{\sigma/\Hn} = -\,\frac{\partial \ }{\partial w}(F,\,f){\ls}_{\sigma/\Hn} = -\frac12f(\vec{P})\,\ .
\end{equation}
Let $p_y$ denote the last component of $\vec{P}$.  Since $p_y = - w$, (\ref{E:SurfInt8}) can be rewritten as
\begin{equation}\label{E:SurfInt9}
\frac{\partial \ }{\partial p_y}(F,\,f){\ls}_{\sigma/\Hn} = \frac12f(\vec{P})\,\,,
\end{equation}
which, after a change of dummy independent variable from $p_y$ to $p'_y$, can be integrated from $p_y$ to $\infty$ to obtain
\begin{equation}\label{E:SurfInt10}
 (F,\,f){\ls}_{\sigma/\Hn} = -\,\frac12\int\limits_{p'_y=p_y}^{\infty}f(\vec{P})\,\,d\,p'_y\ .
\end{equation}

  This last expression can be used to analyze downward continuation, either analytically or numerically.  Consider the $\mathbb{R}^3$ half-space case.  First observe that $(\ell_k,\,{\ell}_{k'}){\ls}_{\sigma/\text{H}^3}$ can be evaluated in closed form from the RHS of (\ref{E:SurfInt10}) so that the associated $\mathbf{T}$ matrix can be easily evaluated.   Second, for specified $f$'s the inner product $(f,\,{\ell}_{k}){\ls}_{\sigma/\text{H}^3}$ can also generally be evaluated either in closed form, in terms of a series expansion or numerically.   For example, in general if a standard $\mathbb{R}^3$ harmonic Fourier integral representation for $f(\vec{X})$ is substituted into the RHS of (\ref{E:SurfInt10}) and the indicated integral is performed then the result seems to imply some sort of rule-of-thumb, which adjusts $\mathbb{R}^3$ downward continuation effects by a wavelength factor.  Alternatively, a general numerical analysis of $\mathbb{R}^3$ downward continuation algorithms could be undertaken by using (\ref{E:SurfInt10}) for various given mathematical representations of $f$ since a numerical evaluation of the RHS of (\ref{E:SurfInt10}) can obviously be performed in this case.  An analysis here might consist of comparing these types of results with results from a standard downward continuation formalism and with results based directly on ${\Phi}_{\infty}$ as specified by (\ref{E:SupNorm}).   The last point is that for a given $f$ and specified set of basis functions, the actual (source) coefficients that yield a minimum of ${\Phi}_{\infty}$ can be found by employing non-linear least squares algorithms.

  It is also perhaps worth noting here that a if a standard Fourier integral harmonic expansion (for $\Sigma$'s of infinite extent) or Fourier series harmonic representation (for $\Sigma$'s of finite extent) is assumed then fits based on (\ref{E:SurfInt10}) can be performed for noisy incomplete data specified on $\Sigma$ (usually either some sort of point or track measurements).  For example, suppose that a collection of noisy point measurements of $f(x,\,y,\,z)$ is available for  $x \in (-L,\,L)$, $y \in (-L,\,L)$ and $z = z_0 > 0$ and a reconstruction of $f(x,\,y,\,0)$ [i.e. $\varphi(x,\,y,\,0) \approx f(x,\,y,\,0)$] is desired where  $x \in (-L',\,L')$, $y \in (-L',\,L')$ with $L' < L$.  Let the associated harmonic Fourier series be written symbolically as (the complete form of the harmonic series can be found in most texts that cover electrostatics):
\begin{equation}\label{E:Fourier}
\mathscr{F}(x,\,y,\,z) \, = \,\sum\ \{a(k_x,\,k_y)\,\cos(k_xx)\cos(k_yy) +\ \text{other sin and cos terms}\}\,e^{-\sqrt{k_x^2+k_y^2}}\,\,,
\end{equation}
so
\begin{equation}\label{E:Fourier2}
-\frac12\int\limits_{z=z_0}^{\infty}\,\mathscr{F}(x,\,y,\,z)\,\,d\,z\, = \,\frac12\,\sum\ \frac{\{a(k_x,\,k_y)\,\,\cos(k_xx)\cos(k_yy) +\ \cdots\}}{\sqrt{k_x^2+k_y^2}} \ .
\end{equation}
Downward continuation can then be carried out by, for example, following the following three step process:
\begin{itemize}
\item[(A)]
Determine the unadjusted expansion coefficients [$a(k_x,\,k_y)$ etc.] by fitting them to the noisy data itself by doing a linear least squares (LLSQ) fit to the data. 
\item[(B)]
Use a frequency domain based procedure to adjust of these coefficients for the presence of noise, so as to obtain a nose compensated surface reconstruction of $f$ for $\Sigma$.  Let $\hat{a}(k_x,\,k_y)$ denote the resulting best unbiased estimates of the new coefficients.
\item[(C)]
Do a fundamental solution downward continuation that is based on inner product evaluations given by the RHS of (\ref{E:Fourier2}), where $\hat{a}(k_x,\,k_y)$ is to be substituted for $a(k_x,\,k_y)$.
\end{itemize}

Several comments are relevant here.  First steps (A) and (B) were successfully implemented by the author in the early 1980's as part of the analysis for an airborne gravity gradiometer survey system project. (The end goal of the project was to obtain surface gravity estimates.)  While the details of the implementation do not concern us here, it is perhaps worth noting that to successfully implement step (A) requires that certain inherent pitfalls be recognized and overcome, but once these underlying issues are recognized the solutions to them are fairly easy to come up with.  \{The main one arises from the interaction of the Gibbs phenomenon with the implied checker-board like periodic repetitions  $\mathscr{F}(x \pm 2L,\,y \pm 2L,\,z) = \mathscr{F}(x,\,y,\,z)$, which produces  unwanted overtones (consider, for example a one dimension Fourier series expansion of a saw-tooth pattern).  Not only must long term biases be removed, but the specified data should be edge tapered in order to remove all of these unwanted edge induced frequency effects.\}  Part of step (C) requires that an appropriate grid and depth for source locations be selected.  It is at this step that some understanding of the interplay between source positions, ``source regularization'' and downward continuation enters.  (In Appendix~B of \cite{DIDACKSII} it was indicated that an understanding of downward continuation is, in some sense, tied to what is known, or is reasonable to assume, in the way of source information or statistics. 
(Various types of ``source regularization'' will be discussed in a subsequent article dealing with inverse source estimation implementation issues.)


\end{document}